\documentclass[aps, prb, preprint, superscriptaddress]{revtex4}
\usepackage{graphicx}
\usepackage[mathcal]{eucal}
\usepackage{multirow}

\begin{document}

\title{Electronic structures of defective BN nanotubes under transverse 
electric fields}

\author{Shuanglin Hu}
\affiliation{Hefei National Laboratory for Physical Sciences at
     Microscale,  University of Science and Technology of
     China, Hefei,  Anhui 230026, People's Republic of China}

\author{Zhenyu Li}
\affiliation{Hefei National Laboratory for Physical Sciences at
     Microscale,  University of Science and Technology of
     China, Hefei,  Anhui 230026, People's Republic of China}

\author{X. C. Zeng}
\affiliation{Department of Chemistry and Nebraska Center for Materials 
     and Nanoscience, University of Nebraska-Lincoln, Lincoln, Nebraska 68588, 
	 USA}

\author{Jinlong Yang}
\thanks{Corresponding author. E-mail: jlyang@ustc.edu.cn}
\affiliation{Hefei National Laboratory for Physical Sciences at
     Microscale,  University of Science and Technology of
     China, Hefei,  Anhui 230026, People's Republic of China}

\date{\today}

\begin{abstract}
We investigate the electronic structures of some defective boron nitride 
nanotubes (BNNTs) under transverse electric fields within density-functional 
theory. (16,0) BNNTs with antisite, carbon substitution, single vacancy, and 
Stone-Wales 5775 defects are studied. 
Under transverse electric fields, the band gaps of the defective BNNTs 
are reduced, similar to the pristine ones. The energy levels of the 
defect states vary with the transverse electric field directions, due to the 
different electrostatic potential shift at the defect sites induced by 
the electric fields. Therefore, besides electronic structure and optical 
property engineering, the transverse electric field can 
be used to identify the defect positions in BNNTs.
\end{abstract}


\maketitle

\section{Introduction}
Boron nitride nanotubes (BNNTs) are wide-gap semiconductors regardless of 
their diameter, chirality, or the number of tube walls.\cite{Rubio} 
This implies that BNNTs own uniform electric and optical 
properties,\cite{Radosavljevic,Lauret,Arenal, Wirtz, Park} and makes BNNTs 
suitable for nanoscale electronic and optical applications.
The external electric field is useful to tune the electronic properties of 
nano materials.\cite{Keeffe, Khoo} 
Khoo \emph{et al.}\cite{Khoo} found that, when transverse electric 
fields are applied, pristine BNNTs show uniform gap closure 
behavior. Transverse electric fields can move 
the conduction band and valence band to two sides of the 
tubes. The band gaps of BNNTs are reduced almost linearly with the increase of 
electric fields. Under the same electric field, they are reduced 
with the increase of the diameter of BNNTs, independent of the 
chirality. The static dielectric response,\cite{Guo} the screened 
polarizabilities,\cite{Wang} and the optical properties of BNNTs under finite 
electric fields \cite{Chen} have also been investigated by theoretical 
calculations. 
On the other hand, the electronic structures and properties of BNNTs can be 
modified by introducing intrinsic defects.\cite{Schmidt, Shevlin, Gou} 
The intrinsic defects can introduce localized states in the band 
gap.\cite{Schmidt} Defects and impurities 
can be used to obtain \emph{n}-type or \emph{p}-type semiconducting 
BNNTs.\cite{Xiang, Zhi} The role of defects and also the role of polarization 
field in BNNTs have been reported.\cite{Zhang}
However, to the best of our knowledge, it is still unclear how 
transverse electric fields affect the electronic structures of BNNTs with 
intrinsic defects. 

In this work, we report the first-principles studies on the electronic 
structures of (16,0) BNNTs with antisite, carbon substitution, single vacancy, 
and Stone-Wales 5775 defects under transverse electric 
fields. Electric fields at three different directions 
perpendicular to the tube axis are applied.

\section{method and model}
Our electronic structure calculations are performed by means of the spin 
polarized density-functional theory implemented in Vienna {\it ab initio} 
simulation package (VASP).\cite{VASP, Kresse} The projector augmented wave 
(PAW)\cite{Blochl} method is used to describe the electron-ion interaction, 
and the cut-off energy is set to 400.0 eV. Perdew-Wang functional\cite{PW91} 
is used for the generalized gradient approximation (GGA). The total energy is 
converged to 10$^{-5}$ eV. The atomic structures are fully relaxed without 
electric field until the forces are less than 0.01 eV/\AA{}. 
The optimized geometrical structures are used to calculate the electronic 
structures under transverse static electric fields. Our test calculations 
on a (16,0) BNNT with B antisite indicate that geometry relaxation under 
electric field gives similar results.

We adopt (16,0) BNNTs with diameter 12.8 \AA{} in our calculations. 
The tube axis is along the $z$ direction. To minimize the interaction 
between adjacent images, we use a large cubic supercell 
(25.000 $\times$ 25.000 $\times$ 4.349 \AA{}$^3$) for the pristine BNNT. 
For BNNTs with antisite and carbon substitution defects, we use a supercell 
with \emph{c}=8.698 \AA{}, twice of the lattice parameter for the 
pristine BNNT. 
And for BNNTs with vacancies and Stone-Wales defects, we use a supercell with 
\emph{c}=17.397 \AA{}.
The special k-points for Brillouin Zone 
integrations are sampled using the Monkhorst-Pack scheme.\cite{Monkhorst} 
For the above three models, 1$\times$1$\times$11,1$\times$1$\times$7, 
and 1$\times$1$\times$3 k-points are chosen, respectively. 

The method of applying static electric field implemented in VASP is in the 
spirit of the work of Neugebauer and Scheffler.\cite{Neugebauer} The 
electrostatic potential decreases along the direction of electric field. 
For convenience, all the defects are put at the positive $x$ side in the 
supercell. The transverse electric field applied from the pristine side to 
the defective side is thus at the positive direction along $x$ 
axis ({$\mathcal{E}_{x}$}). And the effects of electric fields applied 
at the positive $y$ direction ({$\mathcal{E}_{y}$}) and negative 
$x$ direction ({$\mathcal{E}_{-x}$}) are also 
studied. The angle between the electric field and the positive $x$ direction 
in the $xy$ plane ($\beta$) can be used to identify the 
direction of the electric field. The values of $\beta$ for 
{$\mathcal{E}_{x}$}, {$\mathcal{E}_{y}$}, and {$\mathcal{E}_{-x}$} are 
0$^{\circ}$, 90$^{\circ}$, and 180$^{\circ}$, respectively. 
For (16,0) BNNTs, a large electric field is required to 
change the band structures significantly. 
The maximum strength of the applied electric fields in this study is 
0.3 V/\AA{}. 
The strength here is enough to induce distinct change of 
electronic properties of the tubes but no electron emission. 
It is expected that similar results should be 
obtained for larger BNNTs under weaker electric fields.

\section{results and discussion}
First, we examine the electronic properties of a pristine (16,0)BNNT under 
electric field. The pristine (16,0) BNNT is a semiconductor with a 
4.47 eV band gap. The conduction band mainly comes from B atoms, and the
valence band mainly comes from N atoms. 
When the electric field increases from 0.0 to 0.1, 0.2, and 0.3 V/\AA{}, 
the band gap decreases from 4.47 eV to 4.04, 3.52, and finally 2.96 eV. 
It changes almost linearly with the increase of electric field strength, 
which is consistent with the work of Khoo \emph{et al.}\cite{Khoo}. The charge 
densities of the conduction band edge (CBE) states are moved along the 
direction of the electric field, to one side of the tube. And the valence 
band edge (VBE) states are moved to the opposite side of the nanotube. 

\begin{figure}[!hbp]
 \includegraphics[width=15cm]{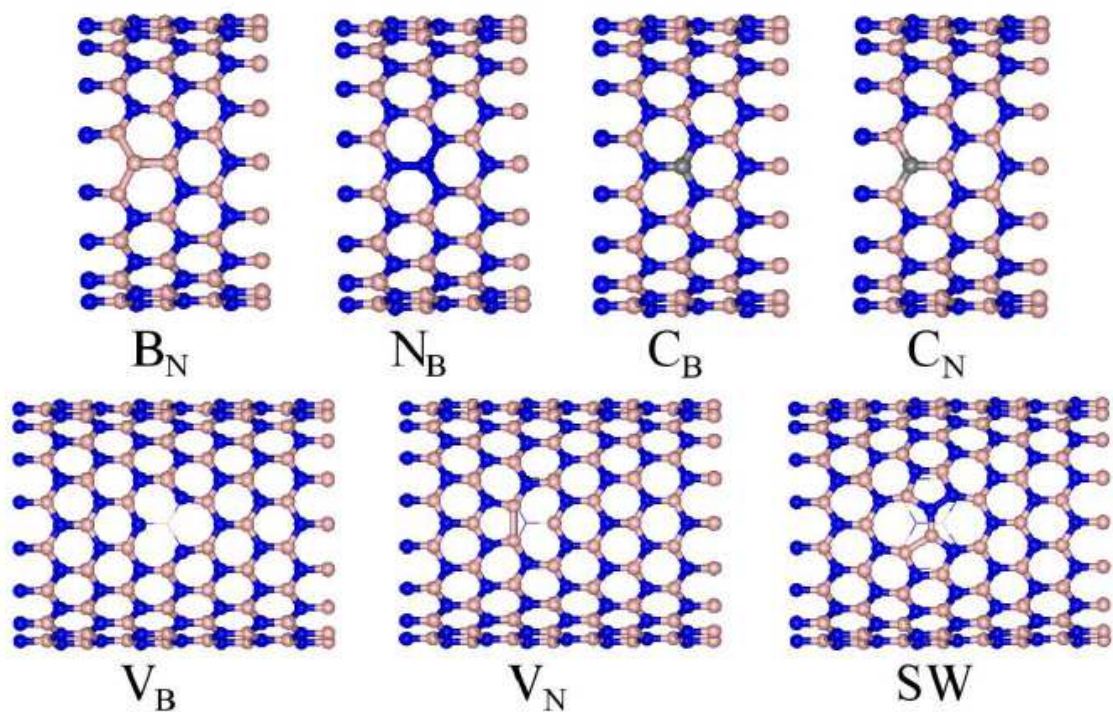}
\caption{The optimized structure of BNNTs with defects.
Pink ball is boron, blue ball is nitrogen, and carbon is in gray.}
\label{fgeo}
\end{figure}

\begin{figure}[!hbp]
 \includegraphics[width=8.5cm]{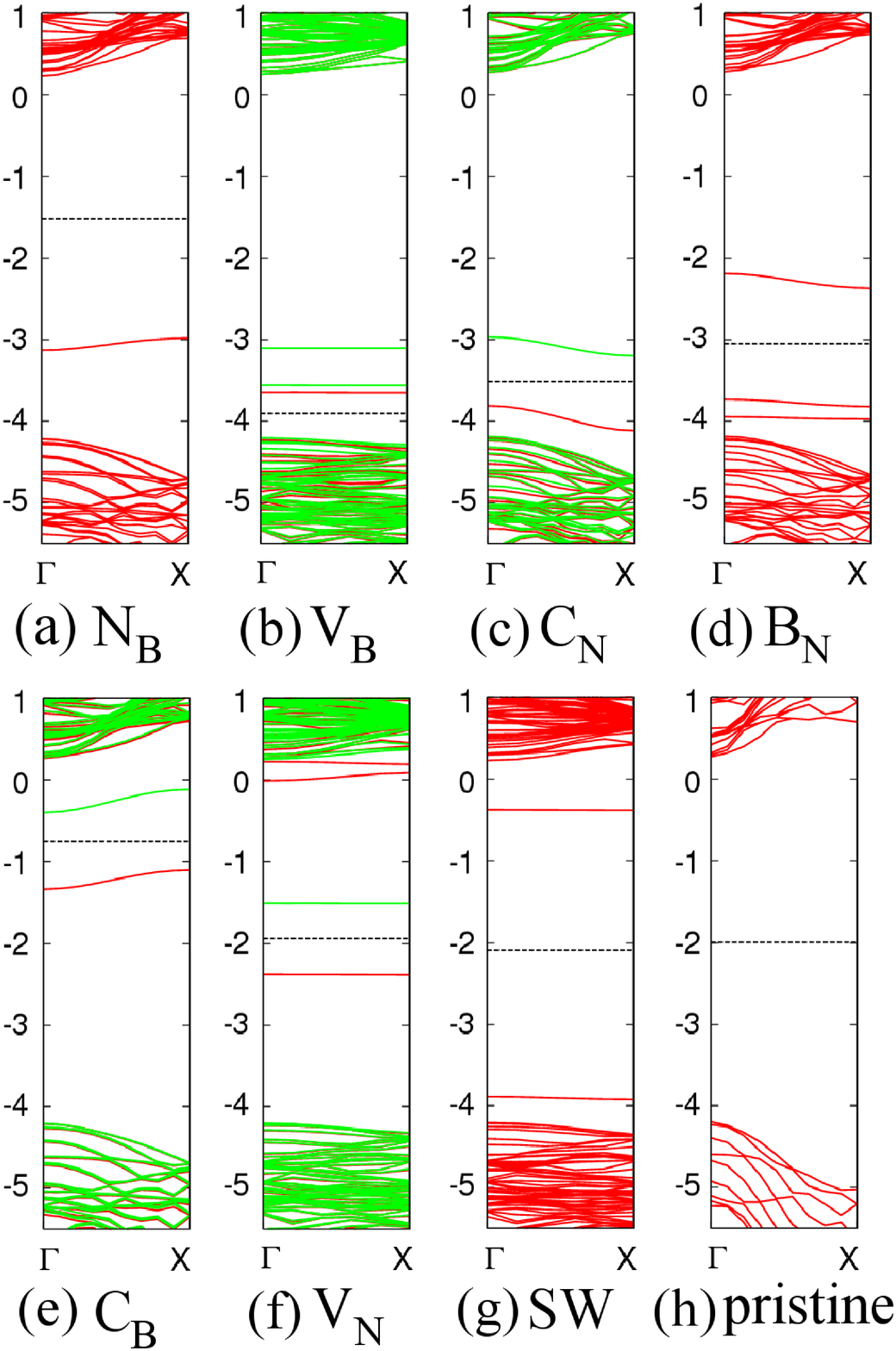}
\caption{Band structures of BNNTs with defects (a) $N_B$, (b) $V_B$, (c) $C_N$, 
(d) $B_N$, (e) $C_B$, (f) $V_N$, (g) SW and (h) pristine BNNTs without 
electric field. The red is the majority spin channel, and green is 
the minority spin channel. The dash line represents the fermi energy.} 
\label{fband}
\end{figure}

\begin{figure}[!hbp]
 \includegraphics[width=8.5cm]{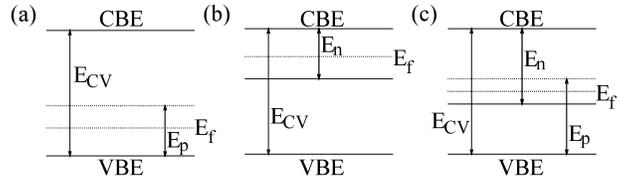}
\caption{The schematic diagram of the energy gaps between the defect states 
and the band edge states. $E_{CV}$ is the gap between the conduction 
band and valence band edge states; $E_{p}$ is the gap between the lowest 
unoccupied defect state and the valence band edge state; and $E_{n}$ is 
the gap between the highest occupied defect state and the conduction band 
edge state. $E_f$ is the fermi energy.}
\label{fdiag}
\end{figure}

\begin{figure}[!hbp]
 \includegraphics[width=15cm]{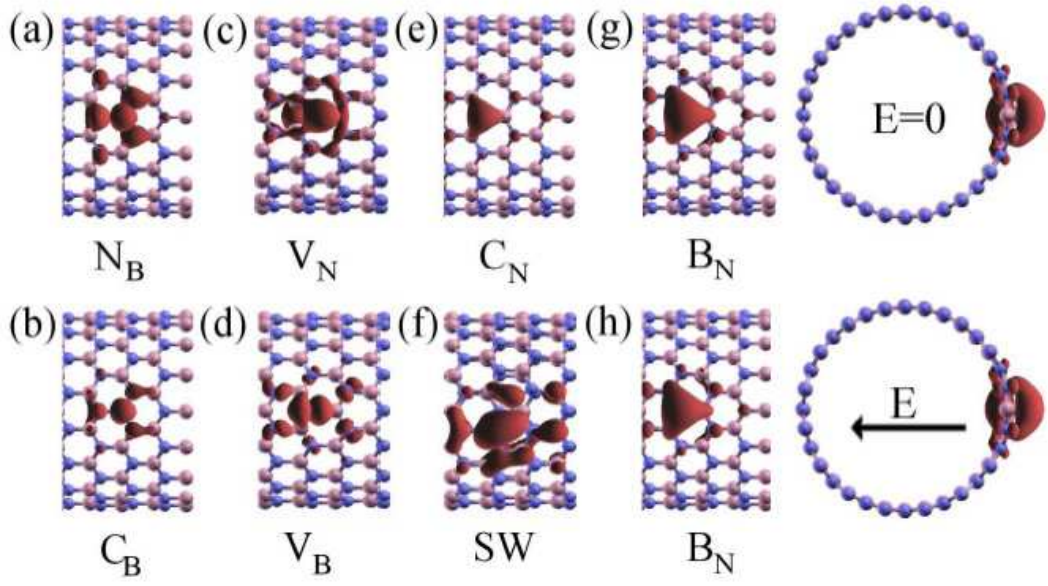}
\caption{The profiles of the $\Gamma$ point charge densities of the highest 
occupied defect states of BNNTs with (a) $N_B$, (b) $C_B$, and (c) $V_N$; 
and the lowest unoccupied defect states of BNNTs with (d) $V_B$, (e) $C_N$, 
and (f) SW under zero electric field. The axial view and 
side view of the lowest unoccupied defect state of BNNT with $B_N$ (g) 
under zero electric field and (h) 0.3 V/\AA{} {$\mathcal{E}_{-x}$} are also 
presented.}
\label{fchg}
\end{figure}

We consider seven kinds of defects on the (16,0) BNNTs: boron 
antisite ($B_N$), nitrogen antisite ($N_B$), carbon 
substitution in a boron site ($C_B$) or a nitrogen site ($C_N$), boron 
vacancy ($V_B$), nitrogen vacancy ($V_N$), and Stone-Wales 5775 (SW). 
The results without electric field are presented first. 
The relaxed geometries are shown in Fig.~\ref{fgeo}. 
The formation energies presented in Table ~\ref{tform} 
are calculated as 
\begin{eqnarray} 
E_{form} & = & E_{binding}[def\text{-}tube]-E_{binding}[ideal\text{-}tube], 
\label{eform} 
\end{eqnarray} 
where the $E_{binding}[ideal\text{-}tube]$ and $E_{binding}[def\text{-}tube]$ 
are binding energies of systems without and with defects, respectively. 
The results are a little larger than those in the (8,0) BNNT. 
\cite{Shevlin, Gou, Wu} 
We also present work function of these BNNTs in the Table ~\ref{tform}.
The band structures of the pristine and defective nanotubes are shown in 
Fig.~\ref{fband}, which are similar with the previous results.\cite{Schmidt} 
The defect $N_B$ introduces only occupied defect 
state, $V_B$ introduces only unoccupied defect states, all the other defects 
introduce both unoccupied and occupied states in the band gap. In (16,0) 
BNNTs with $C_N$ and $B_N$, the unoccupied defect states are close to the VBE, 
but farther than that in the $V_B$ case. 
While in the tube with $C_B$, the occupied defect states are near to the CBE. 
The defect states in BNNTs with $V_N$ and SW are deep in the band gap. 
The band structures of these systems can be sketched schematically in 
Fig.~\ref{fdiag}, we define the energy difference between the CBE and the 
VBE as $E_{CV}$;  the one between 
the lowest unoccupied defect state and the VBE as $E_{p}$;  and the one 
between the highest occupied defect state and the CBE as $E_{n}$. 
From Table ~\ref{tgap}, 
we can see that, under zero electric field, the introduction of defects 
does not change the $E_{CV}$. As for the charge densities of the defect 
states, they are localized near the defect sites. 
It is shown from the profiles of the defect state charge densities 
in Fig.~\ref{fchg}. 
For BNNTs with carbon substitution and single vacancy defects, there are 
also local magnetic moments in these systems (see Table ~\ref{tmag}). 


For the defective (16,0) BNNTs, we study the effects of the 0.3 V/\AA{} 
electric fields at different directions. The $E_{CV}$ in the defective 
systems are all reduced by the electric fields, with slight dependence on 
the field directions. In the narrowed band gap, the energy levels  
of the defect states shift with the field directions remarkably, 
independent of whether the defect states are occupied or not. The 
{$\mathcal{E}_{x}$} lower the energy levels of the defect states. The 
{$\mathcal{E}_{-x}$} uplift the energy of the defect states. The energy levels 
of the defect 
states are between the above two cases, when {$\mathcal{E}_{y}$} is applied. 
As shown in Table ~\ref{tgap}, when electric fields are applied, 
with the increase of the angle $\beta$, the relative movement of the 
defect states and the band edge states makes $E_{p}$ increase and 
$E_{n}$ decrease. In some defective systems, the lowest unoccupied defect 
state can be pushed into the conduction band by the electric fields, it 
makes the absence of the $E_{p}$ in some cases in the Table ~\ref{tgap}. 
And the highest occupied defect state can be also pushed into the valence 
band.
The localized charge densities of 
these defects keep almost unchanged under electric fields. In 
Fig.~\ref{fchg}, as an example, the profiles of the charge density of 
the lowest unoccupied defect state of the BNNT with $B_N$ defect without 
and with 0.3 V/\AA{} {$\mathcal{E}_{-x}$} are shown. 

Although the energy levels of defect states move almost linearly with the 
increase of $\beta$ in all cases, it affects the concentration of carriers 
differently, depending on the kind of defects. For BNNTs 
with $V_B$ and $C_N$ defects, the 0.3 V/\AA{} {$\mathcal{E}_{x}$} 
results in a small $E_{p}$ and enhances the \emph{p}-type conductivity. 
The 0.3 V/\AA{} {$\mathcal{E}_{-x}$} reduces the $E_{n}$ and enhances 
the \emph{n}-type conductivity most for the BNNT with 
$C_B$ defect. In other situations, the $E_{p}$ and $E_{n}$ are maybe still too 
large to improve the conductivity significantly. However, we can use the 
change of conductivity under electric fields with different directions 
to identify the defect positions. 
If we change the directions of the transverse electric fields and measure 
the conductivity, for BNNTs with \emph{p}-type carriers, when the electric 
field is applied from the pristine side to the defective side, we would 
get the max conductance. For BNNTs with \emph{n}-type carriers, we can get 
the max conductance at the reversed field direction. 

For the BNNTs with carbon substitution and single vacancy defects, the local 
magnetic moments show slight dependence on the field directions 
(see Table ~\ref{tmag}), it can be related to the change of $E_{p}$ 
or $E_{n}$. Especially, in the BNNT with $V_B$ defect, when 
the electric fields are applied, with the increase of $\beta$, 
the magnetic moment is reduced. 
As shown in Fig.~\ref{fband}(b), in the band gap, there is one unoccupied 
defect state in the majority spin channel, and two in the minority 
spin channel. The reduction of $E_{p}$ can increase the occupation 
probability on the lowest unoccupied defect state in the majority spin 
channel, so the total local spin can be increased, and the local magnetic 
moment can thus be enhanced. Other cases can be understood in a similar way. 
When the $E_{p}$ and $E_{n}$ are not so small, the magnetic moment changes 
little with the electric field. 

\begin{figure}[!hbp]
\includegraphics[width=8.5cm]{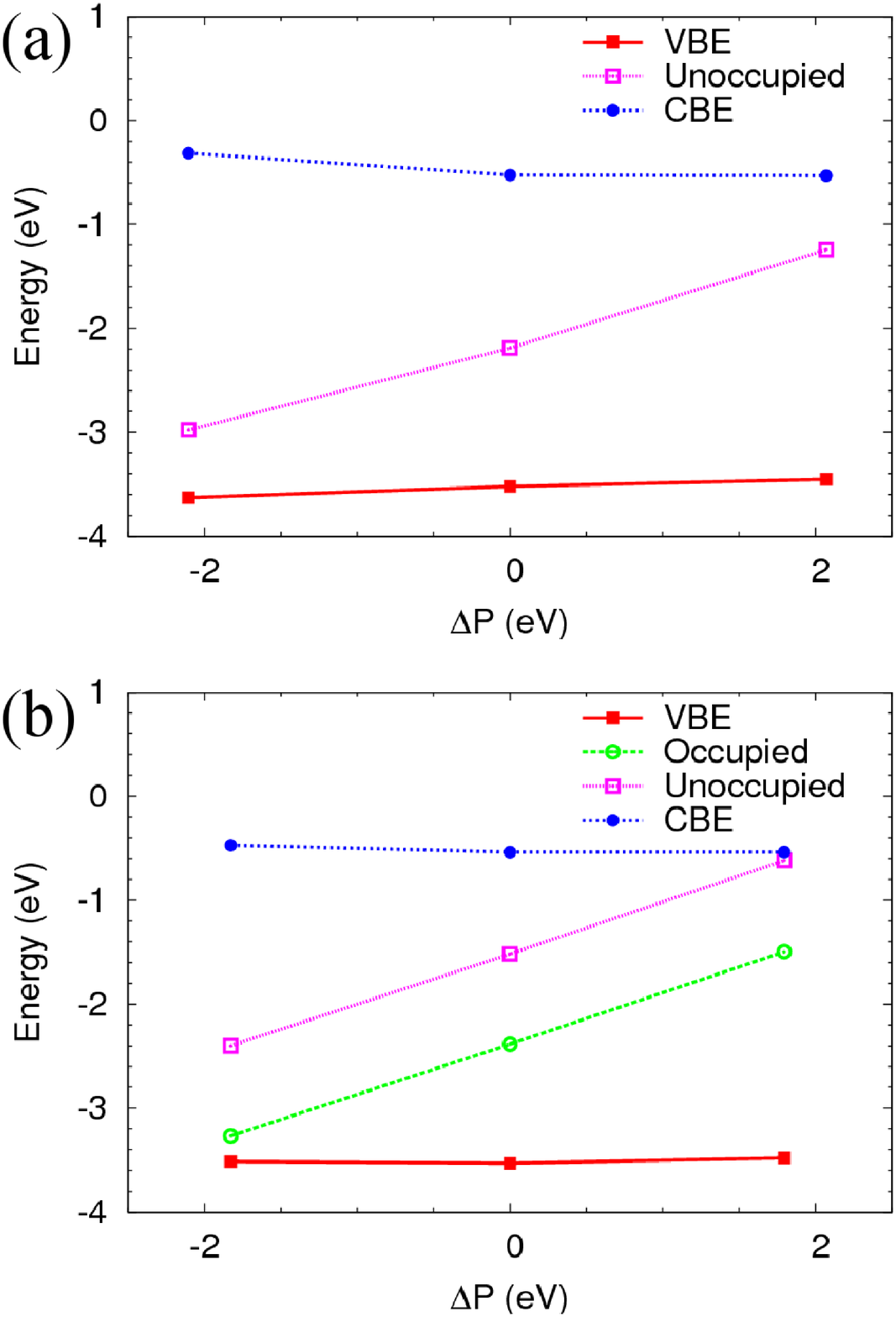}
\caption{The energy levels of the valence band edge (VBE) 
state, the occupied or unoccupied defect states, and the conduction band 
edge (CBE) state of BNNTs with (a) $B_N$ and (b) $V_N$ defects. The abscissa 
is the the shift of electrostatic potential at the defect sites induced 
by electric fields, calculated with Eq.~\ref{eshift}. 
The strength of applied electric fields is 0.3 V/\AA{}.} 
\label{fdef}
\end{figure}

To understand the behavior of defect states under electric fields, we present 
an explanation based on the field induced electrostatic potential shift. 
When {$\mathcal{E}_{x}$} is applied, the defect is at the low potential side. 
While the field is reversed, the defect becomes at a relative high potential. 
So the electrostatic potential at the defect site can be related to $\beta$. 
If we take the center of the BNNTs as origin, at which the electrostatic 
potential is set to zero, the 
potential induced by electric fields at defect sites can be expressed as 
\begin{eqnarray} 
\Delta{}P & = & -|\mathcal{E}| \times x \times cos(\beta), 
\label{eshift} 
\end{eqnarray}
in which $|\mathcal{E}|$ is the strength of the transverse electric field 
(0.3 V/\AA{}), $x$ is the $x$ coordinate of the defect site.
As shown in Fig.~\ref{fdef}, we examine the energy levels of the defect 
states and the band edge states with the $\Delta{}P$ induced by electric 
fields, taking BNNTs with $B_N$ and $V_N$ defects as examples. 
The energy levels of the localized defect states are raised 
significantly with the increase of electrostatic potential at the defect 
sites. However, the levels of CBE and VBE are little affected compared to 
those of the defect states. 
The behavior of $E_{n}$ and 
$E_{p}$ under electric fields thus can be ascribed to the change of 
electrostatic potential at defect sites.
And it can be expected that, stronger {$\mathcal{E}_{x}$} would 
bring down the electrostatic potential at the defect sites further, 
and reduce $E_{p}$ more. The reversed stronger electric fields would 
reduce $E_{n}$ similarly. 

\begin{figure}[!hbp]
 \includegraphics[width=8.5cm]{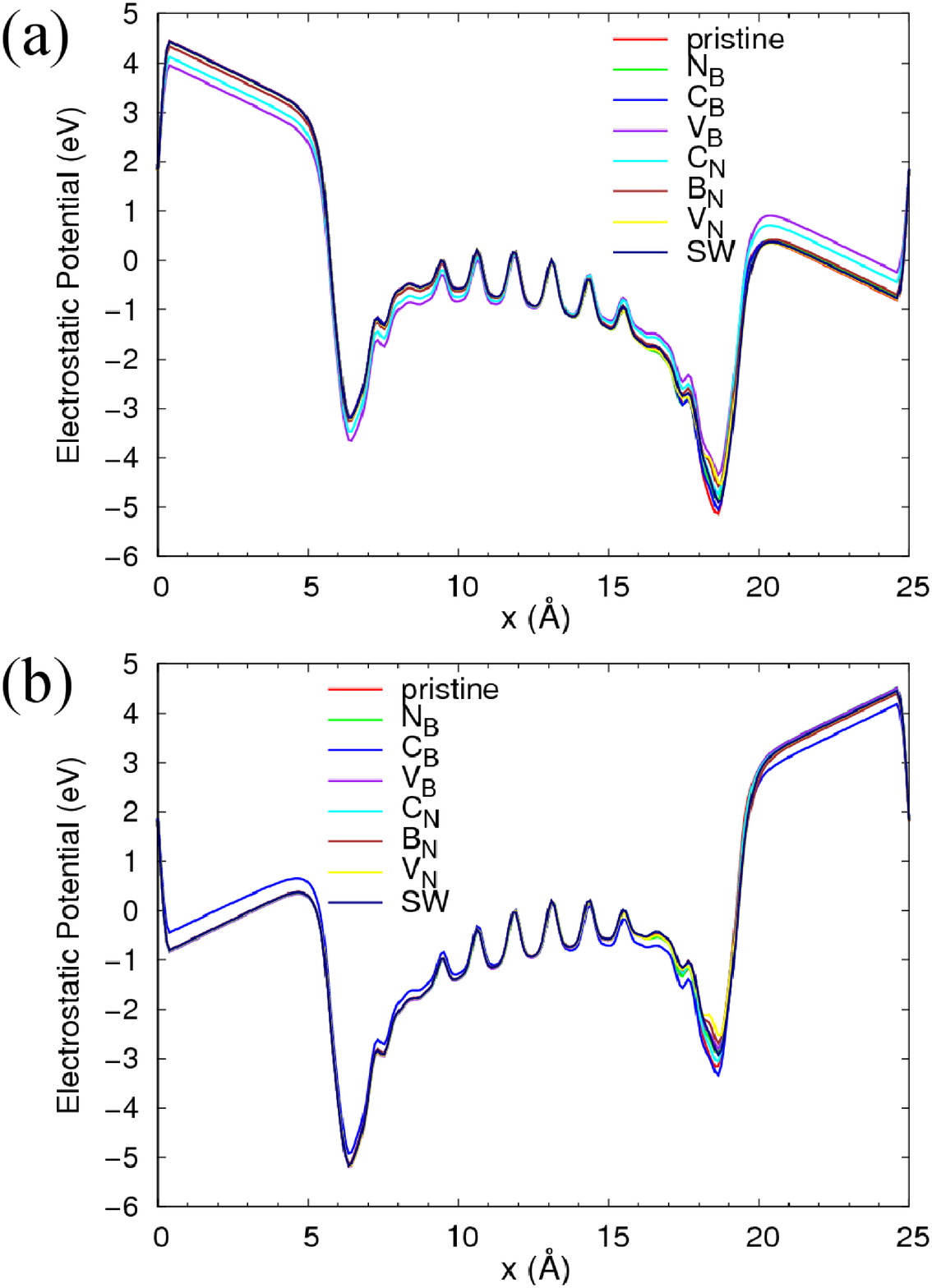}
\caption{The profiles of $yz$-plane-averaged electrostatic 
potentials of pristine and seven defective BNNTs under 
(a) {$\mathcal{E}_{x}$} and (b) {$\mathcal{E}_{-x}$}.}
\label{fpot}
\end{figure}

In most situations, $E_{CV}$ is reduced to the value similar to the pristine 
BNNT under the electric field. But in some cases, the $E_{CV}$ 
is unexpectedly large. Those are BNNTs with $V_B$, $C_N$, and $B_N$ under 0.3 
V/\AA{} {$\mathcal{E}_{x}$} and BNNTs with $C_B$ under 0.3 V/\AA{} 
{$\mathcal{E}_{-x}$}. The large $E_{CV}$ is related to the 
corresponding small $E_{p}$ or $E_{n}$ in these cases. 
The small $E_{p}$ or $E_{n}$ enhances the metallicity of that system, 
and thus increases the screening. With the stronger screening, 
the reduction of electrostatic potential along the direction 
of electric field is slower, therefore, $E_{CV}$ reduction is smaller. 
It can be seen from the profiles of 
$yz$-plane-averaged electrostatic potential under 0.3 V/\AA{} 
{$\mathcal{E}_{x}$} and {$\mathcal{E}_{-x}$}, as shown in Fig.~\ref{fpot}. 

From the above results, we can see that, the electric fields can adjust the 
electronic structures of the defective BNNTs in a different way from pristine 
BNNTs. The transverse electric fields narrow the band gaps, move the 
defect states near to the VBE or CBE, and can change the conductive and 
optical properties smoothly. 
Under the same strength of electric fields, the quantities of 
$E_{p}$ and $E_{n}$ change monotonously with the increase of the angle 
$\beta$. Compared to $E_{p}$ and $E_{n}$, the $E_{CV}$ changes little with 
$\beta$. So transverse electric fields can create shallow acceptor or 
donor states, depending on the kind of defects and the field 
directions. 
For one kind of defective 
BNNTs, one can measure the conductivity under different direction of 
electric fields, to identify the positions of the defects. The change of 
$E_{p}$ and $E_{n}$ under electric fields can also affect the local magnetic 
moments in the BNNTs with single vacancy and carbon substitution defects. 
The charge densities of the localized defect states are all almost unaffected 
by the electric fields. Besides changing their conductivity, 
the electric field engineering of the electronic structures of the 
defective BNNTs can also modify their optical properties.
These external field tunnable properties facilitate application of defective 
BNNTs as nano devices.


\section{conclusion}
In summary, we study the electronic structures of defective BNNTs under 
transverse electric fields. The band gaps of these BNNTs can be reduced by 
electric field like that in pristine BNNTs. The relative positions of 
the defect states in the band gap are moved when the transverse 
electric field direction is changed. The charge densities of 
localized defect states are not affected by the electric field. 
The behavior of electronic structures can be ascribed to the difference 
of the electrostatic potential 
induced by electric fields at the defect sites. Our results indicate that 
defective BNNTs can be used as electric field controlled nano electronic or 
optical devices.

\section*{ACKNOWLEDGMENTS}
This work is partially supported by the National Natural Science
Foundation of China (50121202, 20533030, 20628304),
by National Key Basic Research Program under Grant No. 2006CB922004,
by the Shanghai Supercomputer Center, the USTC-HP HPC project, and the SCCAS.

\clearpage

\begin{table}[!hbp]
\caption{The formation energies ($E_{form}$) and work function ($\Phi$) of 
seven kinds of defects on the wall of the (16,0) BN nanotubes.}\label{tform}
\begin{tabular}{lp{0.3em}ccccccc}
\hline
\hline
         & & $B_N$ & $N_B$ & $C_B$ & $C_N$ & $V_B$ & $V_N$ & SW  \\
\hline
$E_{form}(eV)$ & & 5.37 & 7.34 & 2.70 & 1.55 & 16.08 & 12.01 & 5.71  \\
$\Phi(eV)$     & & 4.86 & 3.91 & 2.58 & 5.33 &  5.74 &  3.76 & 4.81  \\
\hline
\hline
 \end{tabular}
\end{table}

\begin{table}[!hbp]
\caption{The values of $E_{CV}$, $E_{p}$, and $E_{n}$ as plotted in 
Fig.~\ref{fdiag} of pristine and seven kinds of defective (16,0) 
BNNTs under zero and 0.3 V/\AA{} electric fields. The 
electric field is applied at positive $x$, $y$, and negative $x$ direction, 
respectively.}\label{tgap}
\begin{tabular}{cp{0.3em}ccccp{0.3em}ccccp{0.3em}cccc}
\hline
\hline
  &  &  \multicolumn{4}{c}{$E_{CV}(eV)$} & & \multicolumn{4}{c}{$E_{p}(eV)$} & & \multicolumn{4}{c}{$E_{n}(eV)$} \\\cline{3-6}\cline{8-11}\cline{13-16}
Electric field  &  & 0 & {$\mathcal{E}_{x}$} & {$\mathcal{E}_{y}$} & {$\mathcal{E}_{-x}$} & & 0 & {$\mathcal{E}_{x}$} & {$\mathcal{E}_{y}$} & {$\mathcal{E}_{-x}$} & & 0 & {$\mathcal{E}_{x}$} & {$\mathcal{E}_{y}$} & {$\mathcal{E}_{-x}$} \\
\hline                                                      
 pristine & & 4.47 & 2.96 & 2.96 & 2.96 & & - & - & - & - & & - & - & - & - \\
 $N_B$ & & 4.44 & 2.94 & 2.98 & 3.02 & & - & - & - & - & & 3.21 & - & 2.42 & 1.51 \\
 $V_B$ & & 4.45 & 4.03 & 3.26 & 2.90 & & 0.55 & 0.17 & 0.25 & 0.66 & & - & - & - & - \\
 $C_N$ & & 4.46 & 3.74 & 3.07 & 2.91 & & 0.99 & 0.23 & 0.44 & 1.16 & & 4.08 & - &  - & 2.41 \\
 $B_N$ & & 4.46 & 3.31 & 3.00 & 2.92 & & 1.81 & 0.48 & 1.16 & 2.03 & & 4.01 & - & - & 2.29 \\
 $C_B$ & & 4.47 & 2.98 & 3.03 & 3.48 & & 3.80 & 2.21 & - & - & & 1.37 & 1.47 & 0.63 & 0.24 \\
 $V_N$ & & 4.47 & 3.04 & 2.99 & 2.94 & & 2.69 & 1.11 & 2.01 & 2.86 & & 2.65 & 2.79 & 1.85 & 0.96 \\
 SW & & 4.43 & 2.85 & 2.96 & 3.01 & & 3.82 & 2.20 & - & - & & 4.12 & - & - & 2.46 \\
\hline
\hline
 \end{tabular}
\end{table}

\begin{table}[!hbp]
\caption{The magnetic moments per cell ($M$) of (16,0) BNNTs with $C_B$, 
$V_B$, $C_N$, and $V_N$ defects under zero and 0.3 V/\AA{} electric 
fields.}\label{tmag}
\begin{tabular}{cp{0.3em}cccc}
\hline
\hline
 $M({\mu}_B)$     & &   0    & {$\mathcal{E}_{x}$} & {$\mathcal{E}_{y}$} & {$\mathcal{E}_{-x}$} \\
\hline                                              
  $C_B$        & & 1.00 & 1.00 & 0.99 & 0.90 \\
  $V_B$        & & 1.04 & 1.41 & 1.27 & 1.02 \\
  $C_N$        & & 0.99 & 0.87 & 0.97 & 0.99 \\
  $V_N$        & & 1.00 & 1.00 & 1.00 & 1.00 \\
\hline
\hline
 \end{tabular}
\end{table}

\clearpage

%
%
%
%

\begin{verbatim}






\end{verbatim}

\begin{verbatim}

\end{verbatim}
\begin{center}
\end{center}

\clearpage

\begin{verbatim}






\end{verbatim}

\begin{verbatim}

\end{verbatim}
\begin{center}
\end{center}

\clearpage

\begin{verbatim}






\end{verbatim}

\begin{verbatim}

\end{verbatim}
\begin{center}
\end{center}

\clearpage

\begin{verbatim}






\end{verbatim}

\begin{verbatim}

\end{verbatim}
\begin{center}
\end{center}

\clearpage

\begin{verbatim}

\end{verbatim}

\clearpage

\begin{verbatim}






\end{verbatim}

\begin{verbatim}

\end{verbatim}
\begin{center}
\end{center}

\end{document}